\documentstyle[aps,prd,epsfig]{revtex}
\begin{document}
\draft
\title{Stability of string defects in models of \\non-Abelian symmetry breaking}
\author{Marcus J. Thatcher \thanks{Email: Marcus.Thatcher@sci.monash.edu.au; Fax: +61 3 9905 3637; Tel: +61 3 9905 3612}
and Michael J. Morgan \thanks{Email: Michael.Morgan@sci.monash.edu.au; Fax: +61 3 9905 3637; Tel: +61 3 9905 3645}}
\address{Department of Physics, Monash University, Clayton 3168, Victoria, Australia}
\date{\today}
\maketitle
\begin{abstract}
In this paper we describe a new type of topological defect,
called a homilia string, which is stabilized via interactions
with the string network. Using analytical and numerical
techniques, we investigate the stability and dynamics of homilia strings,
and show that they can form stable electroweak strings.
In $SU(2)\times U(1)$ models of symmetry breaking
the intersection of two homilia strings is identified with a sphaleron.
Due to repulsive forces, the homilia strings separate, resulting in sphaleron
annihilation. It is shown that electroweak homilia string loops cannot stabilize 
as vortons, which circumvents the adverse cosmological problems associated with stable 
loops. The consequences for GUT scale homilia strings are also discussed.
\end{abstract}

\section{Introduction}
The inflation model is widely accepted as the standard cosmological model
which describes the Universe from $10^{-43}$ seconds after its creation.
However, standard cosmology presents many unresolved problems, including
the origin of large scale structure of the Universe, the origin of density
fluctuations, which initiate galaxy formation, and the mechanism for baryon
asymmetry, which leads to the production of more matter than antimatter. It
has been conjectured that many of these problems can be resolved by
incorporating topological defects with inflation \cite{vilenkin94}. The most
popular defect model exploits the presence of cosmic strings, which are
predicted by modern Grand Unified Theories (GUTs) to have formed as the early
Universe passed through successive phase transitions \cite{kibble76}.

Around $10^{-35}$ seconds after the Big Bang, the Universe
passed through a phase transition corresponding to symmetry breaking of the
GUT force to distinct strong and electroweak forces. At this
time an extensive GUT cosmic string network is believed to have formed \cite{kibble76}.
The subsequent evolution of this network introduces density fluctuations into
an otherwise homogeneous Universe, which are conjectured to
have lead to large scale structure in the observable Universe \cite{vilenkin85,stebbins87,vachaspati91}.
However, cosmic strings are not topologically stable in simple non-Abelian models,
and therefore are not an inevitable prediction
of all GUT models. Cosmic strings are only stable in models which involve discrete
symmetry breaking \cite{davis97} (e.g., $SU(5) \times Z_{2}$), or when a $U(1)$
generator is gauged to form semi-local strings \cite{vachaspati91b}.
Although cosmic strings provide a mechanism for introducing
fluctuations, they also pose significant problems for cosmology.
Many GUT models predict that a cosmic
string has an internal structure, which allows it to become
superconducting \cite{witten85,everett88}.
Superconducting cosmic string loops can stabilize to vortons \cite{davis88},
which would interfere with primordial nucleosynthesis \cite{brandenberger96}.
In addition to predicting the existence of cosmic strings, GUTs also predict
large numbers of magnetic monopoles \cite{thooft74} which are not observed \cite{preskill79}.
This is often referred to as the monopole problem. However, in models of $SU(2)$
symmetry breaking, sphalerons form instead of monopoles \cite{klinkhamer84}.
Sphalerons are of interest since they are capable of baryon and lepton number violating
processes \cite{leader96}.

In this paper we show that it is possible to construct a cosmologically
significant defect in $SU(2)\times U(1)$ models, that behaves locally like
a string, but is a topologically stable network texture. The local stability
is a consequence of the interactions between adjacent string defects.
Since this new defect is stabilized by the surrounding string network,
we introduce the name homilia\footnote{From the classical Greek meaning `communion'.}
string to differentiate it from cosmic strings.
Sphalerons are identified with the intersection of two homilia strings, and the separation of
the strings results in the annihilation of the sphaleron. Furthermore, 
electroweak homilia string loops cannot stabilize as bosonic or fermonic 
vortons, which circumvents the adverse cosmological problems of stable loops.
Homilia strings can be generalized to $SU(N)\times U(1)$ models of symmetry breaking.
GUT scale homilia strings will induce
fluctuations in the cosmic microwave background (CMB), which are characterized by
greater power at smaller angular scales (in the anisotropy angular power spectrum),
compared to the spectrum generated by cosmic strings.

This paper is organized as follows.
In Sec.\ \ref{sec-lagrangian} we introduce the Lagrangian for
$SU(2)\times U(1)$ symmetry breaking and describe the numerical techniques 
used to analyze homilia strings.
Sec.\ \ref{sec-stable} discusses homilia string stability
and Sec.\ \ref{sec-vortex} examines the homilia string vortex solution.
The dynamics of homilia strings are investigated in Sec.\ \ref{sec-force}. 
The string dynamics are utilized in Sec.\ \ref{sec-cosmo} to highlight the cosmological ramifications 
of homilia strings, including sphaleron decay, network evolution
and vorton instability. The main
results of the paper are summarized in Sec.\ \ref{sec-conclusion}.

\section{A new model of nonsymmetric strings}
\label{sec-homilia}
Both nonsymmetric and cosmic strings are described by the homotopy
group $\pi_{1}$ (i.e., a non-contractible $S^{1}$ circle). However, nonsymmetric
strings \cite{axenides97} differ from cosmic strings since they do not
constitute a line of false vacuum. Within the core of a nonsymmetric
string the Higgs field rotates in isospin space to reconcile an undefined
phase. Usually string defects in non-Abelian models are not topologically
stable because the vacuum symmetry group is larger than the
circle $S^{1}$ that constitutes the string defect. This implies that the
non-simply connected region can be continuously deformed into a simply connected
region, removing the topological defect. However, it is
possible to construct a special type of nonsymmetric string (i.e.,
a homilia string), that is stable in models of $SU(2)\times U(1)$ symmetry breaking.
The stability of homilia strings is related to the topology of
the string network, not to the topology of an individual string. Since a
homilia string network cannot be deformed continuously to the vacuum, it
is topologically stable even in non-Abelian models.

\subsection{Lagrangian and numerical techniques}
\label{sec-lagrangian}
We introduce homilia strings in the context of a
model of spontaneous $SU(2)\times U(1)$ symmetry breaking. The results are
readily generalized to describe homilia strings in non-Abelian $SU(N)\times U(1)$
models. The Lagrangian which describes $SU(2)\times U(1)$ symmetry breaking
is written as (with $c=\hbar=1$)
\begin{equation}
\label{eq-lagrangian}
{\cal L}=(D^{\mu}\Phi)^{\dagger}D_{\mu}\Phi-\frac{1}{4}\mbox{Tr}(F^{\mu\nu}F_{\mu\nu})-\frac{1}{4}\lambda\left(\Phi^{\dagger}\Phi-\eta^{2}\right)^{2},
\end{equation}
where $\Phi$ is an isodoublet, $D_{\mu}=\partial_{\mu}-iqA_{\mu}$,
$F_{\mu\nu}=\partial_{\mu}A_{\nu}-\partial_{\nu}A_{\mu}+iq\left[A_{\mu},A_{\nu}\right]$,
$A_{\mu}=A^{a}_{\mu}\sigma_{a}$ and $\sigma_{a}$ are the Pauli spin matrices ($a=1,2,3$).
To describe $SU(2)\times U(1)$ we also write $\sigma^{0} = \frac{1}{2}I$, where the
associated gauge field is $A^{0}_{\mu}$. The equations of motion are obtained by
varying the Lagrangian (\ref{eq-lagrangian}) with respect to $\Phi^{\dagger}$
and $A^{\mu}$, respectively
\begin{mathletters}
\label{eq-eom}
\begin{eqnarray}
D^{\mu}D_{\mu}\Phi - \frac{1}{2}\lambda\Phi\left(|\Phi|^{2}-\eta^{2}\right) & = & 0 \\
\mbox{Tr}\left(\sigma^{a\,2}\right)\left\{\partial^{\mu}F_{\mu\nu}^{a}-q f^{a}_{bc}A^{b\,\mu}F^{c}_{\mu\nu}\right\}-iq\left\{\left(D_{\nu}\Phi\right)^{\dagger}\sigma^{a}\Phi-\left(\sigma^{a}\Phi\right)^{\dagger}D_{\nu}\Phi\right\} & = & 0,
\end{eqnarray}
\end{mathletters}
where $f_{abc}$ are the structure constants, which for $SU(2)$ are components of 
the fully skew-symmetric tensor $\epsilon_{abc}$.
To facilitate the numerical simulations, the number of parameters in the model is reduced via the
rescaling, $x\rightarrow \eta^{-1} (\lambda/2)^{-1/2} x$, $\Phi \rightarrow \eta \Phi$,
$A_{\mu} \rightarrow \eta A_{\mu}$ and $q\rightarrow e (\lambda/2)^{1/2}$.
This results in the equations of motion (\ref{eq-eom}) having one independent parameter $e$, i.e.,
\begin{mathletters}
\label{eq-eome}
\begin{eqnarray}
D^{\mu}D_{\mu}\Phi - \Phi\left(|\Phi|^{2}-\eta^{2}\right) & = & 0 \\
\mbox{Tr}\left(\sigma^{a\,2}\right)\left\{\partial^{\mu}F_{\mu\nu}^{a}-e f^{a}_{bc}A^{b\,\mu}F^{c}_{\mu\nu}\right\}-ie\left\{\left(D_{\nu}\Phi\right)^{\dagger}\sigma^{a}\Phi-\left(\sigma^{a}\Phi\right)^{\dagger}D_{\nu}\Phi\right\} & = & 0,
\end{eqnarray}
\end{mathletters}
where $D_{\mu}=\partial_{\mu}-ieA_{\mu}$ and $F_{\mu\nu}=\partial_{\mu}A_{\nu}-\partial_{\nu}A_{\mu}+ie\left[A_{\mu},A_{\nu}\right]$.
For convenience we set the parameter $e=1$ in all numerical calculations,
except where stated otherwise.

To analyze the stability of homilia strings
we first calculate the homilia string vortex solution in the
Lorentz gauge, using an iterative relaxation algorithm. This is based on a finite difference
approximation, with appropriate two-point boundary conditions. The
gauge invariant properties of the homilia string are exploited in the numerical simulations 
to identify the formation of homilia strings. We numerically
simulate homilia strings in three dimensions by evolving the fields
using an explicit finite difference
scheme in the temporal gauge ($A_{t}=0$). The latter simulations were performed on a
$100\times 100\times 100$ lattice, with time step $\Delta t=0.1$ and
lattice spacing $\Delta x=0.4$. The stability of a homilia string is related to
interactions with the surrounding network; consequently,
periodic boundary conditions were imposed on the magnitude of the fields
to account for the string network outside the simulation domain. The 
gradient of the phase field was constrained to zero at the boundaries,
since the string configurations described in this paper
exhibited discontinuous phase windings across the boundaries.
The numerical model is modified for simulations of string formation and
is this discussed further in Sec.\ \ref{sec-force}.
Simulations on different lattice sizes were conducted to ensure
that the results are independent of the size of the lattice. The output was generated
by determining which lattice squares contained phase
twists (i.e., $2\pi$ phase rotations). The lattice square which
contained the twist was marked with a light or dark grey dot,
depending on the type of homilia string that was detected. The dots were
then displayed on a three-dimensional plot. To verify that these objects are
homilia strings we use the vortex solution and plots
of the field magnitude, $|\Phi|$; this shows deviations from the vacuum state that are characteristic of homilia strings.
These deviations are associated with a non-zero energy density and hence
homilia strings are not gauge dependent objects.

\subsection{Homilia string stability}
\label{sec-stable}
\begin{figure}
\centerline{\epsfig{file=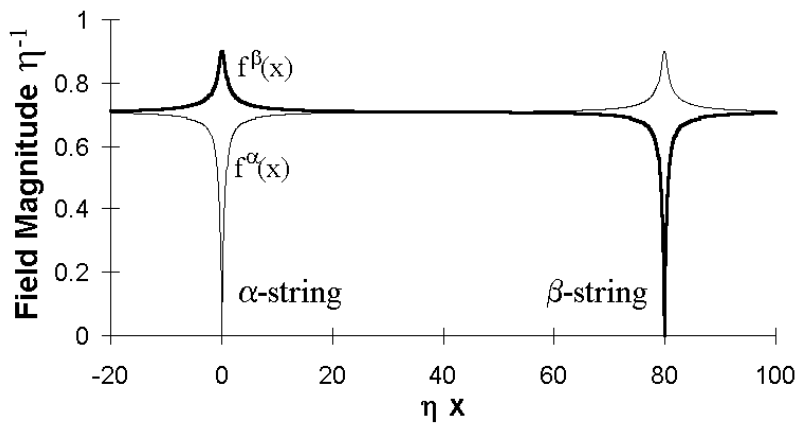,width=.6\textwidth}}
\caption{A cross-section of the field configuration ($e=10$) for an homilia $\alpha$-string
(at $x=0$) adjacent
to a homilia $\beta$-string (at $x=D$). Distances and magnitudes of the
fields have been scaled by $\eta$, as described in Sec.\ \ref{sec-lagrangian}.
This rescaling is adopted for all figures in this paper.
The presence of two types of homilia
strings imposes boundary conditions on the fields $f^{\alpha}(x)$ and $f^{\beta}(x)$,
which result in stable homilia strings. Gauge fields are
not plotted.}
\label{fig-vortextwoa}
\end{figure}

We begin our discussion with an explanation of the homilia string stability
mechanism. The equations of motion and vortex solution are
discussed in Sec.\ \ref{sec-vortex}.
The stability of an individual homilia string is a consequence of the boundary
conditions imposed on a string in the homilia string network.
In models involving $SU(2)\times U(1)$ symmetry breaking, we denote
a homilia $\alpha$-string by the field configuration
\begin{equation}
\label{eq-ss}
\Phi^{\alpha}=\left(\begin{array}{c} f^{\alpha}(r_{1})e^{i\theta_{1}} \\ f^{\beta}(r_{1})e^{i c_{1}} \end{array}\right),
\end{equation}
where $\theta_{1}$ is the topologically non-trivial phase winding for the
$\alpha$-string, centered at some position ${\bf p}$ ($r_{1}=0$), and $c_{1}$ is a constant.
The ansatz (\ref{eq-ss}) has been utilized previously (see e.g., \cite{axenides97}
and \cite{hindmarsh92}). However, our analysis differs from earlier work
since we are interested in the behaviour of Eq.\ (\ref{eq-ss}) when the
homilia string is part of a string network.
The $\alpha$-string is prevented from deforming into
a simply connected region by the presence of a $\beta$-string.
An homilia $\beta$-string is described in $SU(2)\times U(1)$ by the field configuration
\begin{equation}
\label{eq-ssb}
\Phi^{\beta}=\left(\begin{array}{c} f^{\alpha}(r_{2})e^{i c_{2}} \\ f^{\beta}(r_{2})e^{i\theta_{2}} \end{array}\right),
\end{equation}
where $c_{2}$ is a constant independent of $c_{1}$ and
$\theta_{2}$ is another non-trivial phase winding for a $\beta$-string
located at ${\bf q}$ ($r_{2}=0$). The $\alpha$-and $\beta$-strings are separated by a distance $D$.
In Fig.\ \ref{fig-vortextwoa} we have plotted a cross-section
of the fields $f^{\alpha}(x)$ and $f^{\beta}(x)$, for an $\alpha$-string (at $x=0$)
adjacent to a $\beta$-string (at $x=D$). In plotting the cross-section of the
fields we have replaced the radial coordinate $r$ by $x$.
To reconcile an undefined phase requires
\begin{mathletters}
\label{eq-bcone}
\begin{eqnarray}
f^{\alpha}(x=0) & = & 0 \\
f^{\beta}(x=D) & = & 0.
\end{eqnarray}
\end{mathletters}
The total energy of the system is minimized when the deviation of the
magnitude, $|\Phi|$, from $\eta$ assumes its smallest value, whence the
system undergoes symmetry breaking at all points. This
requires $|\Phi(x)|\neq0$, and imposes the conditions
\begin{mathletters}
\label{eq-bctwo}
\begin{eqnarray}
\label{eq-bctwoa}
f^{\alpha}(x=D) & \neq & 0 \\
f^{\beta}(x=0) & \neq & 0.
\end{eqnarray}
\end{mathletters}
The boundary conditions for an $\alpha$-string require
$f^{\alpha}(x)$ to be zero at the center of the $\alpha$-string and
non-zero at the center of the $\beta$-string. If we surround an
$\alpha$-string with $\beta$-strings, then $f^{\alpha}(x)$ cannot be zero
everywhere. Since $f^{\alpha}(x)$ cannot be reduced to zero globally, the
topological defect cannot be deformed to a simply
connected vacuum and the $\alpha$-string is stable.
A similar argument shows that a $\beta$-string is stable in a network of
$\alpha$-strings.
It is important to emphasize that $\alpha$-and $\beta$-strings cannot
exist in isolation and are not independent objects. 

Consider now the situation where an $\alpha$-string has a single $\beta$-string adjacent
to it. In this case, although $f^{\alpha}(x)$ is prevented from going to zero
on one side of the $\alpha$-string, it can go to zero on the other side, and hence the
$\alpha$-string can unwind. However, by placing an $\alpha$-string
in a homilia string network, consisting of both $\alpha$-and $\beta$-strings,
we have 
\begin{equation}
f^{\alpha}(r=D(\theta))\neq0,
\end{equation}
for any given point on the $\alpha$-string. Here $D(\theta)$ is the angular
dependant shortest distance from the $\alpha$-string segment
to a $\beta$-string; $r$ and $\theta$ denote the radial co-ordinates of a point in a plane
taken through the network that intersects the $\alpha$-string
in question. For an arbitrary
direction there will be a $\beta$-string located near the $\alpha$-string at some distance $D(\theta)$,
whence $f^{\alpha}(r)$ is non-zero. Because $f^{\alpha}(r)$
is non-zero at a distance $D(\theta)$ from the $\alpha$-string for all $\theta$, the
homilia $\alpha$-string is a stable defect which cannot unwind.
A similar argument is valid for $\beta$-strings, which leads us to conclude that $\alpha$-and
$\beta$-strings are stable in an $SU(2)\times U(1)$
homilia string network which is unbounded (i.e., an infinite or periodic
network). In such a network all the homilia strings
will be surrounded by homilia strings of the other type.
This situation is accounted for in numerical simulations by imposing periodic boundary conditions
on the magnitude of the fields, which prevents homilia strings
at the edge of the lattice from unwinding.
As long as the entire homilia network
cannot be deformed continuously to the vacuum state individual
homilia strings in the network are stable.

If we assume that the `Mexican hat' potential describing
symmetry breaking is symmetric, and the
fluctuations initiating the phase transition are random, then after
a phase transition there will be phase twists in both components of the $\Phi$ isodoublet.
The interaction between these phase twists leads directly to field configurations
which are described by homilia strings. The phase twists are prevented from
unwinding because each defect makes it energetically unfavourable for
adjacent defects to unwind. The defects are topologically
stable in the network, provided the network consists of the
homilia string types permitted by the symmetry breaking group
(e.g., $\alpha$ and $\beta$ strings). The homilia
string network constitutes a texture defect which locally behaves like
a string. Homilia strings differ from non-topological cosmic strings \cite{saffin97},
since the magnitude of the field, $|\Phi|$, is non-zero at the the core of
the homilia string (i.e., the symmetry is unbroken). Therefore homilia strings
are not associated with a false vacuum.
Homilia strings require the $\Phi$ multiplet components to be complex, and hence
they only exist in models such as $SU(N)\times U(1)$.

\subsection{Homilia vortex solution}
\label{sec-vortex}
To describe a homilia string in a model of $SU(2)\times U(1)$ symmetry
breaking we write the $\Phi$-field in cylindrical form, i.e.,
\begin{equation}
\label{eq-nonsymaz}
\Phi=\left(\begin{array}{c}
\chi^{\alpha} \\ \chi^{\beta} \end{array}\right) = \left(\begin{array}{c}
f^{\alpha}(r)\exp(i n\theta) \\ f^{\beta}(r)\exp(i\kappa^{\beta}) \end{array}\right),
\end{equation}
$\chi^{k}$ ($k=\alpha,\beta$) are complex fields with magnitude $f^{k}(r)$
and phase $\kappa^{k}(\theta)$.
To describe the homilia string we have chosen $\kappa^{\alpha}=n \theta$
(where $n$ indicates the winding number) and $\kappa^{\beta}$ to be constant in the vicinity
of the defect. It is important to emphasize that we have chosen a particular orientation of the
Higgs field, so that the homilia strings separate out into $\chi^{\alpha}$ and
$\chi^{\beta}$ components. In this section we determine the vortex
solution for this particular orientation of the Higgs field in isospin space
(i.e., the `homilia gauge'), for which the two homilia string types separate
into separate components of the $\Phi$ isodoublet. We are then able to
determine gauge invariant properties of homilia strings, which enables us to identify homilia
strings in the three-dimensional network simulations.
In the homilia gauge, it is apparent that the phase winding of
the $\alpha$-and $\beta$-strings do not couple through the potential.

To determine the vortex solution, consider an $\alpha$-string described by
\begin{equation}
\label{eq-defalpha}
\Phi=\left(\begin{array}{c} f^{\alpha}(r)e^{i\theta} \\ f^{\beta}(r) \end{array}\right)=e^{i\tau^{\alpha}\theta}\left(\begin{array}{c}f^{\alpha}(r) \\ f^{\beta}(r) \end{array}\right),
\end{equation}
where we have set the phase of the $\chi^{\beta}$-field to zero,
and $\tau^{\alpha}$ is the generator for the $\alpha$-string in the homilia gauge.
This generator is written as
\begin{equation}
\tau^{\alpha} = \left[\begin{array}{cc} 1 & 0 \\ 0 & 0 \end{array}\right].
\end{equation}
The properties of $\tau^{\alpha}$ in Eq.\ (\ref{eq-defalpha})
can be determined from a Taylor expansion, utilizing $(\tau^{\alpha})^{2}=\tau^{\alpha}$.
Similarly, a $\beta$-string generator can be written as
\begin{equation}
\tau^{\beta} = \left[\begin{array}{cc} 0 & 0 \\ 0 & 1 \end{array}\right].
\end{equation}
With this particular orientation of the $\Phi$-field (i.e., the homilia gauge),
the generators are Abelian
\begin{equation}
\left[\tau^{\alpha},\tau^{\beta}\right]=0.
\end{equation}

To accommodate gauge fields in the vortex solution we define
\begin{eqnarray}
A_{\theta}(r,\theta) & = & \frac{n}{e r}\left[b(r)\tau^{\alpha}+c(r)\left(\cos\theta\sigma^{1}-\sin\theta\sigma^{2}\right)\right] \nonumber \\
\label{eq-hgauge}
& = & \frac{1}{2}\left[\begin{array}{cc} 2 b(r) & c(r) e^{i\theta} \\ c(r)e^{-i\theta} & 0 \end{array}\right],
\end{eqnarray}
where $b(r)$ and $c(r)$ are real fields and $\sigma^{1}$, $\sigma^{2}$ and 
$\sigma^{3}$ are the Pauli spin matrices defined by
\begin{mathletters}
\begin{eqnarray}
\sigma^{1} & = & \left[\begin{array}{cc} 0 & 1/2 \\ 1/2 & 0 \end{array}\right] \\
\sigma^{2} & = & \left[\begin{array}{cc} 0 & -i/2 \\ i/2 & 0 \end{array}\right] \\
\sigma^{3} & = & \left[\begin{array}{cc} 1/2 & 0 \\ 0 & -1/2 \end{array}\right],
\end{eqnarray}
\end{mathletters}
with $\sigma^{0}=\frac{1}{2}I$.

The equations governing the homilia string vortex solution are obtained by
substituting the ansatz [Eqs.\ (\ref{eq-nonsymaz}) and (\ref{eq-hgauge})] into the equations of motion
(\ref{eq-eome}), to obtain (in the Lorentz gauge):
\begin{mathletters}
\label{eq-gaugevortex}
\begin{eqnarray}
(f^{\alpha}(r))''+\frac{(f^{\alpha}(r))'}{r}-\frac{n^{2}f^{\alpha}(r)}{r^{2}}\left(b(r)-1\right)^{2}-\frac{n^{2}c(r)}{2 r^{2}}\left(\frac{f^{\alpha}(r)c(r)}{2}+f^{\beta}(r)b(r)\right) & & \nonumber \\
-f^{\alpha} \left(|\Phi|^2-\eta^{2}\right) & = & 0 \\
(f^{\beta}(r))''+\frac{(f^{\beta}(r))'}{r}+\frac{n^{2} f^{\beta}(r) c(r)^{2}}{4r^{2}}+\frac{f^{\alpha}(r) c(r)}{2 r^{2}}\left(1-b(r)\right)- f^{\beta} \left(|\Phi|^2-\eta^{2}\right) & = & 0 \\
b(r)''-\frac{b(r)'}{r} - e^{2} \left[2 (f^{\alpha}(r))^{2}\left(b(r)-1\right)+f^{\alpha}(r)f^{\beta}(r)c(r)\right] & = & 0 \\
c(r)''-\frac{c(r)'}{r} - e^{2} \left[2f^{\alpha}(r)f^{\beta}(r)\left(b(r)-1\right)+|\Phi|^{2}c(r)\right] & = & 0. 
\end{eqnarray}
\end{mathletters}
where a dash denotes differentiation with respect to $r$ and
\begin{equation}
|\Phi|^{2}=\sum_{k=\alpha,\beta} (f^{k}(r))^{2}.
\end{equation}
In the case of an $SU(2)\times U(1)$ homilia $\alpha$-string we want the
boundary conditions to describe a neighboring $\beta$-string at $r=D(\theta)$.
The boundary conditions (\ref{eq-bcone}) and (\ref{eq-bctwo}) imply that the
magnitudes $f^{\alpha}(x)$ and $f^{\beta}(x)$ are equal at a distance mid-way between
the $\alpha$-string and the nearest $\beta$-string (see Fig.\ \ref{fig-vortextwoa}).
A cylindrically symmetric solution can be obtained by approximating the distribution 
of the surrounding $\beta$-strings, at the average distance $r=\zeta/2$ halfway 
between the segments of homilia $\alpha$-string and the surrounding $\beta$-strings,
with $f^{\alpha}(r=\zeta/2)\approx f^{\beta}(r=\zeta/2)$.
Assuming the strings do not overlap, this enables us to write the cylindrically symmetric boundary 
conditions for the approximate vortex solution of the $\alpha$-string as
\begin{mathletters}
\label{eq-sutwobc}
\begin{eqnarray}
f^{\alpha}(r=0)=0 \\
\left[\frac{d(f^{\beta}(r))}{dr}\right]_{r=0}=0 \\
f^{\alpha}(r= \zeta/2) \approx f^{\beta}(r=\zeta/2).
\end{eqnarray}
\end{mathletters}
The boundary conditions (\ref{eq-sutwobc}) differ from those employed in
reference \cite{axenides97} for conventional nonsymmetric strings,
due to the orientation of the $\Phi$-field in isospin space (i.e., the choice of the homilia gauge).
The boundary conditions for $b(r)$ and $c(r)$ in Eqs.\ (\ref{eq-hgauge}),
consistent with Eqs.\ (\ref{eq-gaugevortex}), are
\begin{mathletters}
\begin{eqnarray}
b(r=0) & = & 0 \\
b(r=\zeta/2) & \approx & 1 \\
c(r=0) & = & 0 \\
c(r=\zeta/2) & \approx & 0.
\end{eqnarray}
\end{mathletters}

A property of the vortex solution (\ref{eq-gaugevortex}) is that
$|\Phi|$ is not equal to $\eta$ everywhere. To demonstrate this, consider
a global vortex solution where $e=0$. The equations
of motion for the case $|\Phi|=\eta$ are
\begin{mathletters}
\begin{eqnarray}
(f^{\alpha}(r))''+\frac{1}{r}(f^{\alpha}(r))'-\frac{n^{2}}{r^{2}}f^{\alpha}(r) & = & 0 \\
(f^{\beta}(r))''+\frac{1}{r}(f^{\beta}(r))' & = & 0.
\end{eqnarray}
\end{mathletters}
The solutions to the fields $f^{\alpha}(r)$ and $f^{\beta}(r)$ are written as
\begin{mathletters}
\label{eq-deviation}
\begin{eqnarray}
f^{\alpha}(r) & = & \frac{A}{r^{n}}+\frac{B r^{n}}{2 n} \\
f^{\beta}(r) & = & C + D\ln(r),
\end{eqnarray}
\end{mathletters}
where $A$, $B$, $C$ and $D$ are constants. The boundary conditions
(\ref{eq-sutwobc}) imply that $A=0$ and
$B\neq0$, and hence the solutions (\ref{eq-deviation})
cannot satisfy $|\Phi|=\eta$ everywhere.
We conclude that the presence of a homilia string forces the $|\Phi|$-field
to depart from the vacuum state.
For homilia strings with gauge fields, the numerical calculations indicate that
the result is still valid at the center of the string, where $b(r)$ and $c(r)$ are small.
This deviation is what gives homilia strings their independent energy density,
and demonstrates that they are not equivalent to the vacuum state. Consequently, homilia
strings have a tension which causes string loops to collapse.
Collapsing loops, together with intercommuting, provide
homilia string networks with an energy loss mechanism in the same
manner as $U(1)$ cosmic string networks \cite{shellard87}. This prevents the homilia
string network from dominating the energy density of the Universe.

\begin{figure}
\centerline{\epsfig{file=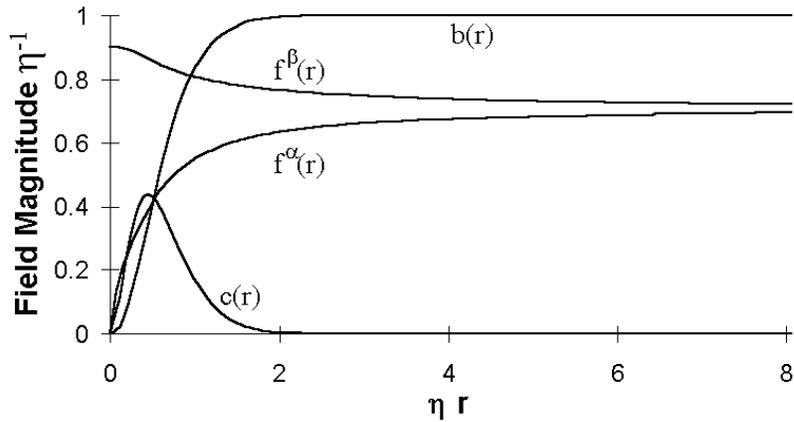,width=.6\textwidth}}
\caption{The vortex solution for a homilia string in
$SU(2)\times U(1)$ ($e=10$). $f^{\alpha}(r)$, $f^{\beta}(r)$, $b(r)$ and $c(r)$ are defined in the text.}
\label{fig-gsvortex}
\end{figure}

\begin{figure}
\centerline{\epsfig{file=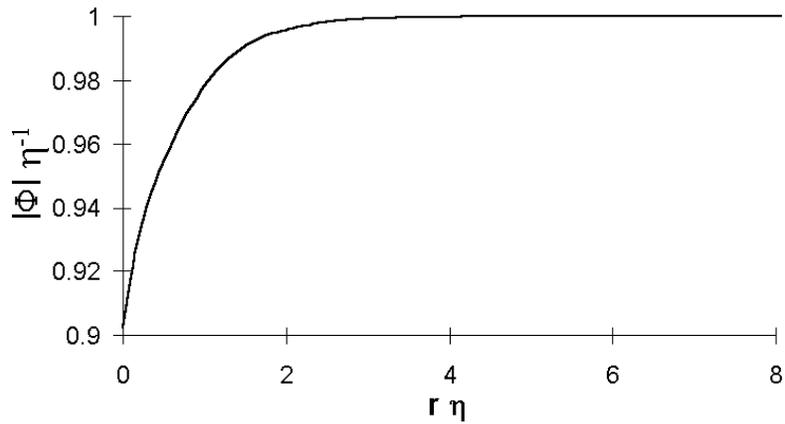,width=.6\textwidth}}
\caption{Plot of the magnitude $|\Phi|$ for the $SU(2)\times U(1)$ homilia string vortex
solution in Fig.\ \ref{fig-gsvortex}. The deviation of $|\Phi|$ from
$\eta$ at the core of the string is evident when gauge fields are included.
This deviation provides a major contribution to the energy per unit length
in a homilia string.}
\label{fig-deviation}
\end{figure}

In Fig.\ \ref{fig-gsvortex}
we have plotted an $SU(2)\times U(1)$ homilia string vortex solution for $e=10$.
The vortex solution was calculated numerically from the equations of motion
(\ref{eq-gaugevortex}), subject to the boundary conditions (\ref{eq-sutwobc}).
The deviation of $|\Phi|$ from $\eta$ at the center of the string
is evident in Fig.\ \ref{fig-deviation}, which shows that a deviation
is present when gauge fields are included.
This deviation results in a non-zero energy density for the homilia string.

Although Eqs.\ (\ref{eq-deviation}) show that there must be a deviation of
the $|\Phi|$-field from the vacuum state, they do not indicate the size
of the deviation.  In the case of global $SU(2)$ symmetry breaking, the
size of the deviation depends on the distance between two homilia strings.
As the average distance between global homilia strings increases, the energy
per unit length of each global homilia string decreases and quickly becomes negligible. However, in the
case of a homilia string interacting with a gauge field (e.g., $SU(2)\times U(1)$),
the gauge field screens the defect in the same way that cosmic string interactions
are screened for large $r$ in local $U(1)$ symmetry breaking models \cite{myers92}.
This means that for local homilia strings the deviation of $|\Phi|$ from the 
vacuum state does not depend on the average separation distance, but rather on $e$.  
Increasing the relative coupling strength results in
the gauge field approaching its vacuum solution more rapidly
with distance from the string core, and increases the deviation of the
$|\Phi|$-field from the vacuum at $r=0$. For $e=1$ the maximum deviation of $|\Phi|$ from
$\eta$ is about $1\%$, but for $e=10$ the deviation is $\sim10\%$.
Estimates of the Higgs boson
mass suggest that $m_{\Phi}>40$ GeV \cite{leader96}, and hence this places an upper
bound on $e$:
\begin{equation}
\label{eq-ewbound}
\frac{m_{W}}{m_{\Phi}}=e < 2,
\end{equation}
where $m_{W}$ denotes the mass of the intermediate vector boson.
If we assume that the dominate contribution to the energy per unit length is from
the potential term, we can establish an upper bound on the electroweak 
($\eta\sim10^{2}$ GeV) homilia string mass per unit length (from the vortex 
solution as)
\begin{equation}
\mu_{H}< 10^{-3}\eta^{2} \approx 10\,\mbox{GeV}^{2} \approx 10^{-7}\,\mbox{g/m}.
\end{equation}
This is several order of magnitude smaller than an `electroweak cosmic string' (i.e., false vacuum string),
which has a mass per unit length of $\sim10^{-4}$ g/m.

\subsection{Interaction forces and homilia string dynamics}
\label{sec-force}
\begin{figure}
\centerline{\epsfig{file=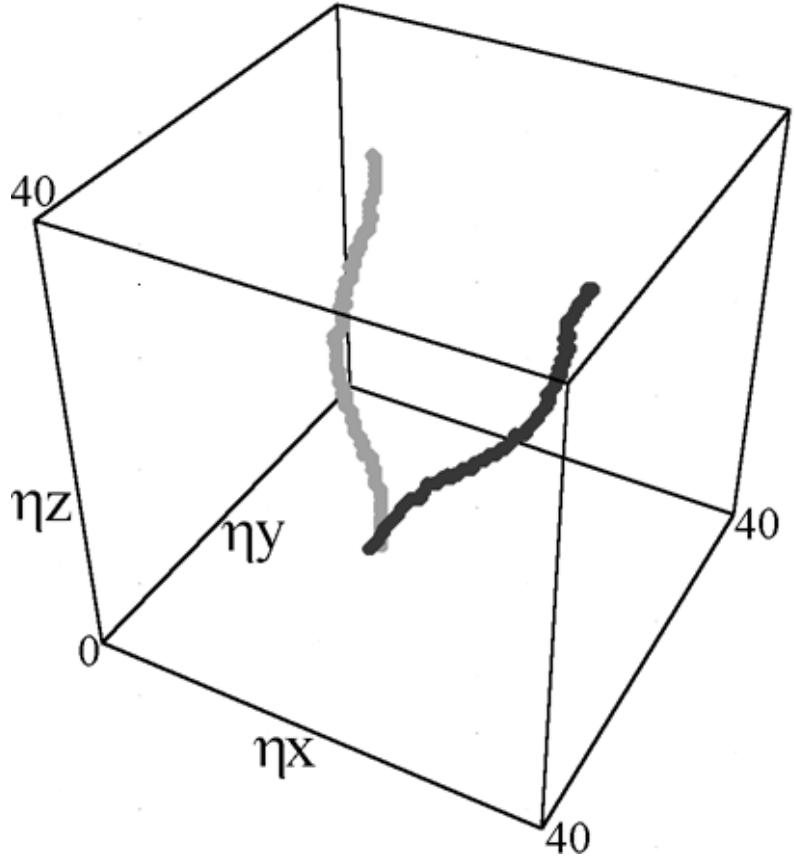,width=.6\textwidth}}
\caption{Two $SU(2)\times U(1)$ homilia strings of different
types ($\alpha$ is light grey and $\beta$ is dark grey),
at $t=12 \eta^{-1}$  ($e=1$). Despite being orientated
at right angles, the two homilia strings are stable in the presence of each other.
The bending of the homilia strings is indicative of the
repulsive interaction force between homilia strings of different types.}
\label{fig-gsstable}
\end{figure}

The stability of homilia strings in three dimensions was investigated for
two homilia strings orientated perpendicular to each other. In Fig.\ \ref{fig-gsstable}, 
we have plotted two (initially stationary) homilia strings in an $SU(2)\times U(1)$ model. 
The three-dimensional simulations were conducted in the temporal gauge ($A_{t}=0$).
Homilia strings were initially wound into the fields using the vortex solution
(modified for the temporal gauge), with
multiple homilia strings, separated by a reasonable distance,
superposed in the homilia gauge using the ansatz
\begin{mathletters}
\begin{eqnarray}
\chi^{i}_{T} & = & 2^{(M-1)/2}\prod_{k=1}^{M} \chi^{i}_{k} \\
A_{T}^{\mu} & = & \sum_{k=1}^{M} A^{\mu}_{k},
\end{eqnarray}
\end{mathletters}
where $\chi^{i}_{k}$ and $A^{\mu}_{k}$ denote the fields of the individual
homilia strings and $M$ is the number of superposed strings. The factor
$2^{(M-1)/2}$ is used to normalize the superposed fields.
To account for the surrounding homilia string network in
Fig.\ \ref{fig-gsstable}, periodic boundary conditions are imposed.
The $\alpha$-string is initially orientated parallel to the $z$-axis and the
$\beta$-string is orientated parallel to the $y$-axis.
After evolving the fields, the $\alpha$-and $\beta$-strings
were found to be stable, suggesting that homilia strings are stable for
arbitrary orientations in three dimensions.

\begin{figure}
\centerline{\epsfig{file=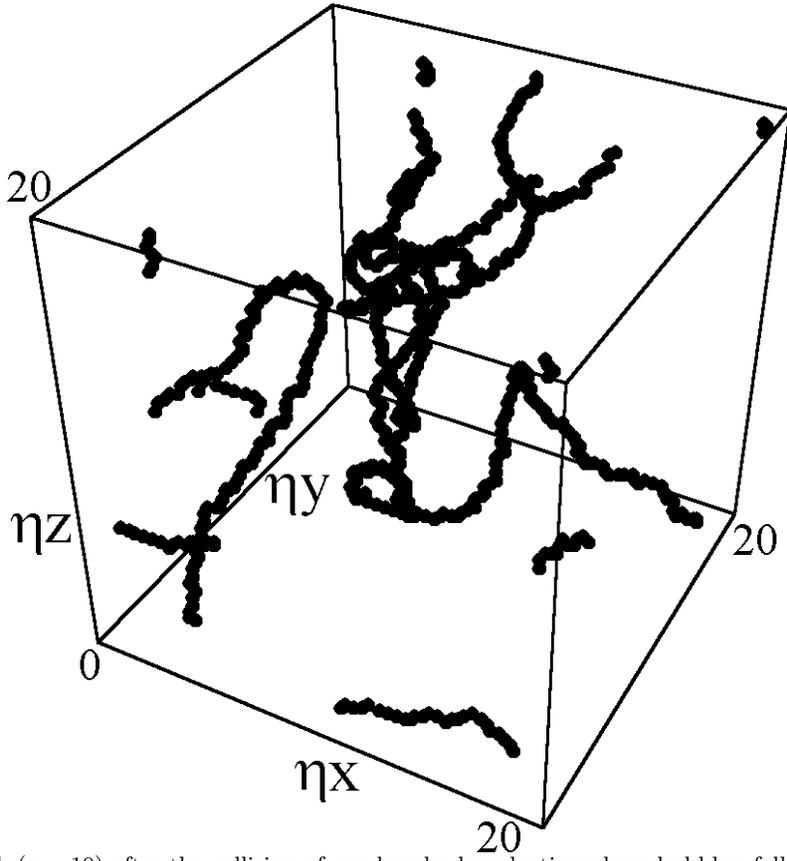,width=.6\textwidth}}
\caption{String network ($e=10$) after the collision of one hundred nucleating
phase bubbles, following an $SU(2)\times U(1)$ first order phase transition ($t=15 \eta^{-1}$).
The resultant network is constructed from homilia $\alpha$-strings 
and $\beta$-strings.  Since we have not chosen the orientation of $\Phi$,
we cannot easily distinguish homilia strings of different types from the
phase fields, and hence both string types are assigned black.}
\label{fig-gsform}
\end{figure}

The stability of homilia strings is examined further
in Fig.\ \ref{fig-gsform}, where we plot the results of a
simulation of one hundred nucleating bubbles of random phase and position,
originating from a $SU(2)\times U(1)$ first order phase transition.
For this simulation, we have set $e=10$, $\Delta t=0.025$ and changed the size of the
simulation to a $50\times50\times50$ cubic lattice.
The boundary conditions were modified to
be periodic in the magnitude and phase of $\Phi$. As a consequence of the violent behaviour
of $|\Phi|$, arising from bubble nucleation and string crossing (see Sec.\ \ref{sec-sphaleron}),
we have added a damping term, $d\dot{\Phi}$,
to the equations of motion (\ref{eq-eome}), where $d$ is the damping factor.
For the simulation depicted in Fig.\ \ref{fig-gsform}, the homilia strings
were evolved relativistically ($d=0$) for $t<8\eta^{-1}$.
For $t\geq 8\eta^{-1}$ the damping factor was increased smoothly to $d=-2$, which
made deviations in $|\Phi|$ from the vacuum, corresponding to homilia 
strings (Fig.\ \ref{fig-deviation}), clearly visible.
Ideally we would choose
the damping term to be as small as possible; however, due to the finite size
of the simulation, if the damping term is too small the string density is
reduced to zero (via intercommuting and loop production), before the
field settles sufficiently for deviations in $|\Phi|$ to be clearly distinguished.
Moreover, in Sec.\ \ref{sec-sphaleron} we show that the intersection of homilia
$\alpha$-and $\beta$-strings are
associated with the formation and annihilation of sphalerons (i.e., points of
false vacuum). Hence, the damping factor must remain non-zero for the entire 
duration of the simulation as the formation and annihilation of sphalerons introduces
fluctuations in $|\Phi|$.

\begin{figure}
\centerline{\epsfig{file=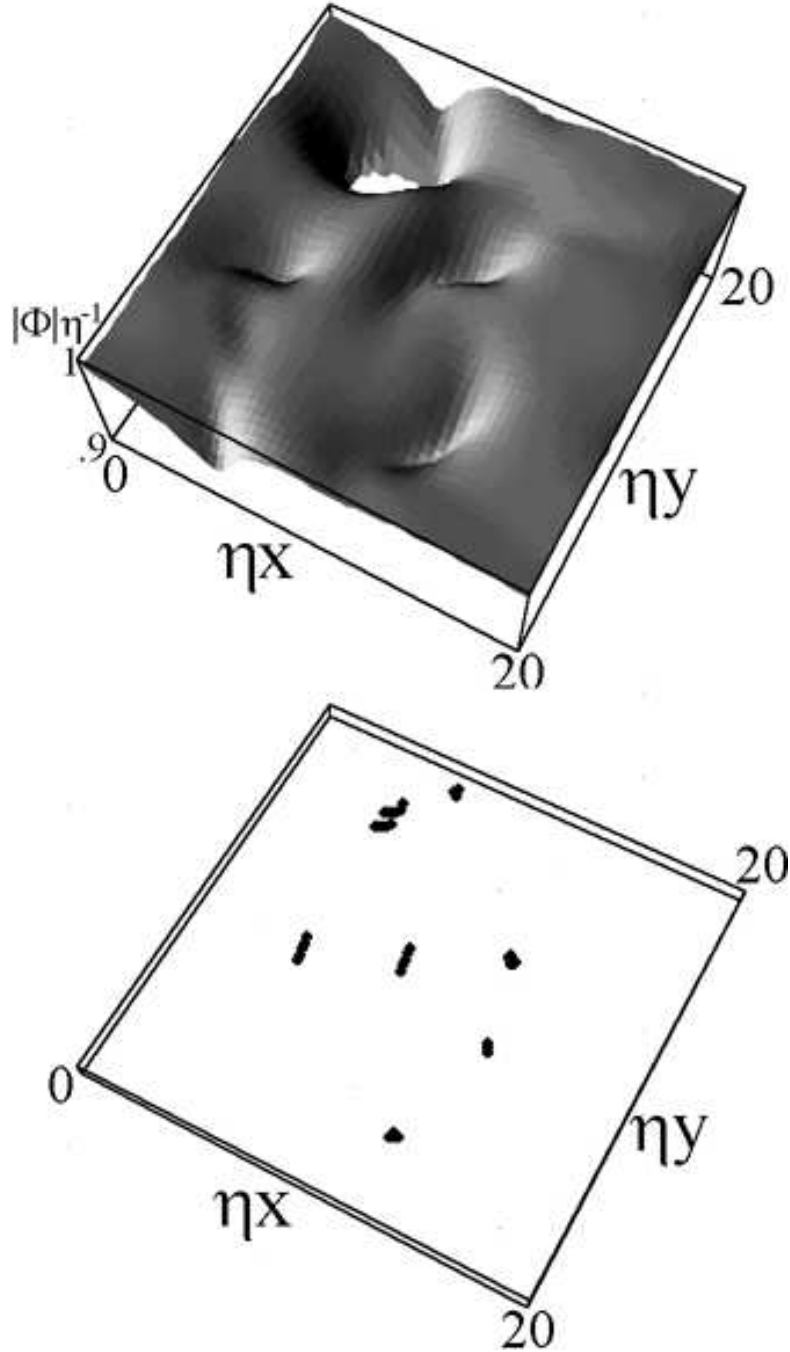,width=.6\textwidth}}
\caption{Plot of $|\Phi|$ ($e=10$) through a constant slice at $z=10\eta^{-1}$
in Fig.\ \ref{fig-gsform}. The positions of the homilia strings, identified
from the points of undefined phase, are also plotted.
The plot of the $|\Phi|$-field is clipped at $|\Phi|=0.9\eta$.
The presence of homilia strings (corresponding the the defect positions) 
is indicated by deviations of $|\Phi|$ from $\eta$ of approximately
$5\% - 7\%$. The deviations are
in reasonable agreement with the prediction of $10\%$ from the vortex solution in Fig.\ \ref{fig-deviation}.
Note that the deviations are generally elongated since
many of the deviations overlap and most string defects
are not perpendicular to the plane of the slice. The large deviation in
the background corresponds to the formation of a sphaleron, due to the intersection of two homilia
strings of different types (see Sec.\ \ref{sec-sphaleron}).}
\label{fig-hsdev}
\end{figure}

The resultant behaviour of the fields after the $SU(2)\times U(1)$ phase transition is plotted in Fig.\ \ref{fig-gsform}, 
which shows the formation of several homilia strings
of type $\alpha$ and $\beta$. Homilia strings are stable even in
a complicated network. In Fig.\ \ref{fig-gsform}, we cannot anticipate the local orientation of
$\Phi$ in isospin space. Consequently, after formation the $\alpha$-and $\beta$-strings will not
separate into distinct components of the $\Phi$ isodoublet,
and hence we cannot trivially distinguish homilia strings of different types.
Therefore, both string types have been shaded black in Fig.\ \ref{fig-gsform}.
Homilia strings are still observed to form when the
damping term is removed, although this makes the interpretation of the behaviour
of $|\Phi|$ ambiguous due to its violent behavior. 
In Fig.\ \ref{fig-hsdev} we have plotted the magnitude of the $\Phi$-field for a constant slice
in Fig.\ \ref{fig-gsform} (i.e., for $z=10\eta^{-1}$). Deviations of $|\Phi|$ from the vacuum
of $5\% - 7\%$, corresponding to homilia strings
are clearly observed. Consequently, homilia strings are associated with
gauge invariant properties (i.e., a deviation in $|\Phi|$ which results
in a non-zero string mass per unit length), and are not gauge dependent objects.
The discrepancy between the observed deviations
of $5\% - 7\%$ and the predicted deviation of $10\%$ from the vortex solution
can be attributed to the non-uniform distribution of the strings. This
results in the cylindrically
symmetric vortex solution (\ref{eq-gaugevortex}) being a poorer approximation 
to the defect field behaviour when the average separation is small.
The deviations are expected to become better approximated by
the vortex solution as the average separation distance increases with time
(see Sec.\ \ref{sec-network}). To analyze
a realistic homilia string network, with large average separation
distances, requires a significant increase of the lattice size in the numerical simulation.
Nevertheless, Fig.\ \ref{fig-hsdev} shows that homilia
strings are associated with gauge invariant properties, in particular a deviation
in $|\Phi|$ which contributes to a non-zero string energy per unit length.

In Fig.\ \ref{fig-gsstable} it is observed that two homilia strings `bend' under the influence of
a repulsive (interaction) force between the strings. Homilia strings which are of different types
(i.e., $\alpha$ and $\beta$ strings) will always experience a repulsive force.
This is because the force is related to the stability of homilia strings,
originating from the formation of a sphaleron when the two strings
intersect (Sec.\ \ref{sec-sphaleron}). The angular interaction force 
on a segment of homilia $\alpha$-string due to a segment of $\beta$-string is calculated from
\begin{equation}
\label{eq-cforce}
\left.F(D)\right|_{\theta=\theta_{0}} =-\frac{\partial}{\partial D}\int_{0}^{D}\epsilon(r,\theta=\theta_{0})\,dr,
\end{equation}
where $\epsilon(r,\theta=\theta_{0})$ is the energy density of the string
along the line $\theta=\theta_{0}$ joining the two string segments;
$D$ represents the separation between the segments.
In the case of an
homilia $\alpha$-string adjacent to a $\beta$-string (Fig.\ \ref{fig-vortextwoa}),
the field $f^{\alpha}(r)$ is zero at the center of the $\alpha$-string and
non-zero at the center of the $\beta$-string. When the $\alpha$-and $\beta$-strings
are in close proximity (for small $D$), the boundary conditions cannot be satisfied
without increasing the energy of the system.
Equation (\ref{eq-cforce}) implies that a increase in the energy corresponds
to a repulsive force. In the case of cosmic strings,
the interaction between strings can be attractive (for a certain choice
of parameters), due to the gauge field overscreening the force. These
strings are referred to as type I \cite{myers92}. However, the overscreening
depends on the non-trivial phase winding of the two cosmic strings being 
described by the same generator (i.e., the
phase fields are the same).
This is not the case for homilia strings of different types,
which can be seen for strings in the homilia gauge (Sec.\ \ref{sec-vortex});
here the $\alpha$-and $\beta$-strings are wound in different $\chi^{k}$-fields
(i.e., $\chi^{\alpha}$ and $\chi^{\beta}$ for $\alpha$-and $\beta$-strings, respectively).
Consequently, the force between homilia strings of different types is always repulsive.

Equation (\ref{eq-cforce}) gives the angular force on a $\alpha$-string segment
due to an adjacent segment of $\beta$-string, i.e.,
\begin{equation}
\left.F(D)\right|_{\theta=\theta_{0}}=-\epsilon(D,\theta=\theta_{0}).
\end{equation}
For $D=0$, the two string segments are superposed to form a point  
of false vacuum, $|\Phi|=0$ (see Sec.\ \ref{sec-sphaleron}). 
This means that the force will be a maximum (as $D\rightarrow0$)
at the maximum deviation of $|\Phi|$ from the vacuum. Moreover, $\epsilon(D=0)$ is
finite, where $\epsilon(D=0)\sim\frac{1}{2}\lambda\eta^{4}$
and consequently the repulsive force is finite. Given sufficient velocity,
homilia strings of different types can collide and pass
through each other. The different types of homilia strings do not
intercommute because the Mexican hat potential for an $SU(2)\times U(1)$
model has the form
\begin{equation}
\label{eq-mh}
V(|\Phi|)=\frac{1}{4}\lambda\left[\{f^{\alpha}(r)\}^{2}+\{f^{\beta}(r)\}^{2}\right]^{2}.
\end{equation}
Only the magnitudes, $f^{\alpha}(r)$ and $f^{\beta}(r)$, couple through
the potential, not the phase fields $\kappa^{k}$ associated with the
two perpendicular homilia string generators.
In a previous paper we argued that intercommuting is
a consequence of phase dynamics \cite{thatcher97}. Since the phase fields of
homilia strings of different types do not interact directly (see Sec.\ \ref{sec-vortex}),
the homilia strings do not intercommute and instead pass through each
other. In the case of a homilia string network,
homilia strings of different types initially become entangled,
which results in a rapid increase in the string tension.
The increased tension compensates for the repulsive force and the homilia strings
pull through each other. Therefore, homilia string networks will only become
entangled for brief intervals of time. This conjecture is supported by the
network formation simulation (Fig.\ \ref{fig-gsform}), where $\alpha$-and
$\beta$-strings are observed to pass through each other in a
collision.

\begin{figure}
\centerline{\epsfig{file=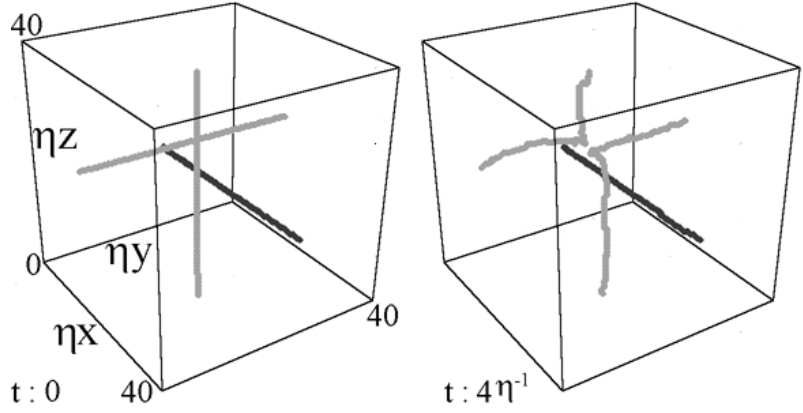,width=.6\textwidth}}
\caption{Intercommuting strings of the same type ($\alpha$-strings
denoted by light grey) with $e=1$.
Since the two homilia $\alpha$-strings
are wound in the same $\kappa^{\alpha}$-phase, the corresponding magnitude $f^{\alpha}(r)$
drops to zero between the strings. This quickly brings the two $\alpha$-strings into
contact, at which point they intercommute. Homilia string networks can
form loops which subsequently collapse and provide the string network
with an energy loss mechanism. The $\beta$-string (dark grey) is introduced
to prevent the
system from globally deforming into the simply connected vacuum state.}
\label{fig-dynamics}
\end{figure}

In order to produce a stable $SU(2)\times U(1)$ homilia string, we require a string of one
type to be surrounded by the other string type (e.g., an $\alpha$-string
surrounded by $\beta$-strings).
However, after formation the homilia string network will be quite complicated
and it is possible that a homilia string will be adjacent to a
homilia string of the same type.
Homilia strings of the same type interact rapidly, as they are unstable and locally
unwind (i.e., $f^{\alpha}(x)\rightarrow 0$ between two strings).
Such a scenario is depicted in Fig.\ \ref{fig-dynamics}, where
two homilia $\alpha$-strings, initially orientated perpendicular to
each other, are observed to be unstable.
The extra homilia $\beta$-string serves to prevent the entire system
from globally deforming to the vacuum. The two $\alpha$-strings display
complicated dynamics, with string segments moving towards each other
and colliding. As the phase twists of the two $\alpha$-strings are
wound in the same $\kappa^{\alpha}$-phase (associated with the homilia $\alpha$-string
generator), the two $\alpha$-strings intercommute and move apart due to string
tension. Therefore, homilia string networks can form
loops, which subsequently collapse due to string tension. This
allows homilia strings to reduce the energy of the string network,
preventing homilia strings from dominating the Universe. This mechanism is
identical to that for $U(1)$ cosmic string networks \cite{shellard87}.

\section{Consequences of homilia strings}
\label{sec-cosmo}
\subsection{Sphaleron annihilation}
\label{sec-sphaleron}
Sphalerons are point defects of false vacuum that form in models of
$SU(2)$ symmetry breaking \cite{klinkhamer84}. However,
a topologically stable string network in $SU(2)\times U(1)$
produces sphalerons in the context of homilia strings.
Consider the intersection between a homilia $\alpha$-and $\beta$-string,
where the $\alpha$-string is parallel to the $z$-axis and the $\beta$-string is
parallel to the $y$-axis and the point of intersection is at $x=y=z=0$. We write the
$\Phi$-field for $SU(2)\times U(1)$ in the form
\begin{equation}
\label{eq-sutwo}
\Phi=\left(\begin{array}{c} \chi^{\alpha} \\ \chi^{\beta} \end{array}\right)=\left(\begin{array}{c} f^{\alpha}e^{i\kappa^{\alpha}} \\ f^{\beta}e^{i\kappa^{\beta}} \end{array}\right),
\end{equation}
where $\chi^{k}=\chi^{k}(x,y,z)$, $f^{k}=f^{k}(x,y,z)$ and $\kappa^{k}=\kappa^{k}(x,y,z)$
($k=\alpha,\beta$). In the homilia gauge (Sec.\ \ref{sec-vortex}), the
intersection of the two homilia strings is determined by the phase winding
\begin{mathletters}
\label{eq-monoconst}
\begin{eqnarray}
\tan(\kappa^{\alpha}) & = & \frac{x}{y} \\
\tan(\kappa^{\beta}) & = & \frac{x}{z}.
\end{eqnarray}
\end{mathletters}
Since $f^{\alpha}=0$ along the $\alpha$-string ($z=0$) and $f^{\beta}=0$ along the
$\beta$-string ($y=0$), at the point of intersection ($x=y=z=0$) we
have $f^{\alpha}=f^{\beta}=0$. This is a point of false vacuum and can be interpreted
as a sphaleron. Since a sphaleron represents the intersection of
an $\alpha$-and $\beta$-string, it is not an independent object.

\begin{figure}
\centerline{\epsfig{file=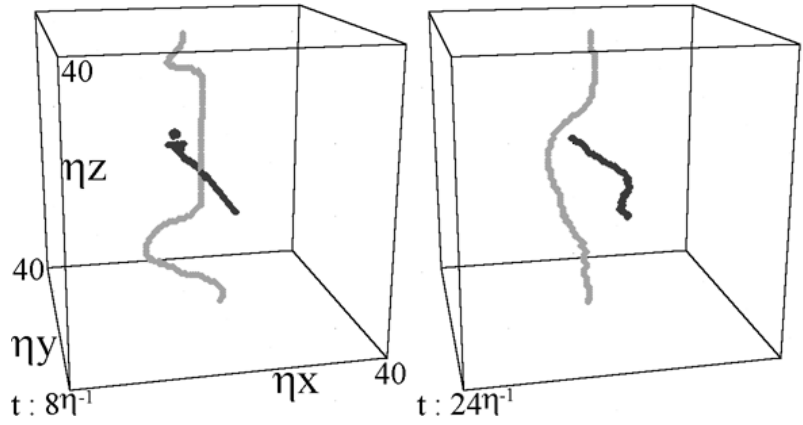,width=.6\textwidth}}
\caption{The decay of an $SU(2)\times U(1)$ sphaleron ($e=1$). 
The sphaleron is constructed from the
intersection of an homilia $\alpha$-string and $\beta$-string (light and dark grey,
respectively). After being perturbed by a kink, the two strings
separate and the sphaleron decays. A small secondary kink propagating
down the $\alpha$-string can be seen at $t=8\eta^{-1}$.}
\label{fig-sphaleron}
\end{figure}

In Sec.\ \ref{sec-force} we argued that homilia strings of different types
exhibit a repulsive force for any choice of parameters. This is because the
energy associated with the sphaleron configuration is much greater than
that of separate homilia strings, which results in a repulsive force.
The repulsive force is finite as the energy of the sphaleron is finite.
Consequently, the sphaleron configuration of two intersecting homilia
strings (\ref{eq-monoconst}) is inherently unstable. Given a perturbation (e.g.,
a kink propagating along one of the strings), the two homilia strings will separate, resulting in sphaleron
decay (see Fig.\ \ref{fig-sphaleron}). Perturbations in the string
network can be generated by a number of mechanisms, including kinks
arising from intercommuting and the collapse of homilia string loops.
Collapsing loops pull the two homilia strings apart annihilating the sphaleron.

Sphalerons which form after a $SU(2)\times U(1)$ phase
transition can be viewed as the points of intersection between homilia
$\alpha$-and $\beta$-strings. As the homilia string network evolves, the
intersecting homilia strings separate, annihilating the points of false
vacuum and the sphalerons decay. To prevent the network from becoming
entangled, homilia strings of different types will sometimes pass through
each other (see Sec.\ \ref{sec-force}). Such events correspond to the rapid formation
and annihilation of sphalerons.
However, since sphalerons are related to the intersections of strings in the
homilia string network, the sphaleron density will decrease as the homilia string
network density decreases (see Sec.\ \ref{sec-network}).

\subsection{Isospin rotations}
\label{sec-isospin}
In the electroweak model, the $SU(2)\times U(1)$ Lie generators
are fixed in isospin space relative to the Higgs vacuum, i.e.,
\begin{equation}
\Phi_{EW}=\eta\left(\begin{array}{c} 0 \\ 1 \end{array}\right).
\end{equation}
Regardless of the actual state, we can rotate $\Phi$ in isospin
space to $\Phi_{EW}$, using a local gauge transformation $U(x)$, i.e.,
\begin{equation}
\Phi_{EW}=U(x)\Phi.
\end{equation}
$U(x)$ is expressed in its most general form by
\begin{equation}
U(x) = V \exp(i{\bf \Theta}(x)\cdot {\bf\sigma}),
\end{equation}
where $V$ is a constant matrix that describes a global gauge transformation, 
${\bf\Theta}(x)$ is a real vector and ${\bf\sigma}$ is a vector corresponding 
to the Pauli spin matrices.

Consider the form of the homilia $\alpha$-string (\ref{eq-defalpha})
at large distances from the string core. In terms of the boundary conditions
(\ref{eq-sutwobc}), when $r\rightarrow\zeta/2$ the $\alpha$-string is given by
\begin{equation}
\Phi^{\alpha}(r\rightarrow\zeta/2)\approx\frac{\eta e^{i\tau^{\alpha}\theta}}{\sqrt{2}}\left(\begin{array}{c} 1 \\ 1 \end{array}\right).
\end{equation}
To describe the electroweak vacuum state at large distances, we introduce a
gauge transformation
\begin{mathletters}
\begin{eqnarray}
\Phi \rightarrow U(x) \Phi \\
A_{\mu} \rightarrow U(x) A_{\mu} U(x)^{-1} - i g^{-1} U(x)^{-1} \partial_{\mu} U(x).
\end{eqnarray}
\end{mathletters}
For the far field of an
homilia $\alpha$-string to be consistent with the electroweak vacuum state,
we must perform a gauge transformation of the form
\begin{eqnarray}
U(\theta) & = & V \exp(i{\bf \Theta(\theta)}\cdot {\bf\sigma}) \nonumber \\
\label{eq-ewtrans}
& = & \frac{1}{\sqrt{2}}\left(\begin{array}{cc} -1 & 1 \\ 1 & 1 \end{array}\right) e^{-i \tau^{\alpha}\theta},
\end{eqnarray}
so that
\begin{equation}
\Phi^{\alpha}(r\rightarrow\zeta/2)\rightarrow U(\theta) \Phi^{\alpha}(r\rightarrow\zeta/2) \rightarrow \Phi_{EW}.
\end{equation}
Note that $U(\theta)$ is not defined at $r=0$, where $\theta$ is undefined.

Although we can locally gauge away the phase variation in $\Phi$, we cannot
gauge away variations in the magnitudes of the $\Phi$ multiplet [i.e,
$f^{\alpha}(r)$ and $f^{\beta}$(r)]. 
In the limit as $r\rightarrow0$, the boundary conditions
require $f^{\alpha}(r\rightarrow0)\rightarrow0$ and $f^{\beta}(r\rightarrow0)\rightarrow \kappa(e)\eta$, where
$\kappa(e)$ is the maximum deviation of $|\Phi|$ from the vacuum.
Therefore, as $r\rightarrow0$, in the vicinity of the $\alpha$-string we find
\begin{equation}
\label{eq-nearrotation}
\Phi^{\alpha}(r\rightarrow0) \rightarrow U(\theta)\Phi^{\alpha}(r\rightarrow0) \rightarrow \frac{\kappa(e)\eta}{\sqrt{2}}\left(\begin{array}{c} 1 \\ 1 \end{array}\right)\neq \Phi_{EW},
\end{equation}
By choosing a gauge so the $\Phi$-field is consistent with the electroweak model
far from the homilia string, 
$\Phi$ must rotate in isospin space away from $\Phi_{EW}$ as $r\rightarrow0$.
A similar analysis applies to $\beta$-strings and consequently homilia
$\alpha$-and $\beta$-strings cannot be gauged away. This rotation has
implications for the superconductivity of homilia strings discussed in
Sec.\ \ref{sec-sc}.

\subsection{Superconducting homilia strings and vortons}
\label{sec-sc}
Cosmic string defects can have an internal structure and become superconducting
\cite{witten85,everett88}.
Superconducting cosmic strings are classified according to whether they have
bosonic, fermonic or non-Abelian gauge supercurrents.
However, superconducting cosmic string loops have adverse cosmological
consequences since they stabilize as vortons \cite{davis88},
which interfere with primordial nucleosynthesis \cite{brandenberger96}.
Homilia strings avoid these problems, since the behaviour
of the supercurrent inside a homilia string is different to
that of a cosmic string. To show this we examine superconductivity in
homilia strings.

Bosonic supercurrents couple to the false vacuum through the potential
\cite{witten85}
\begin{equation}
V(|\Phi|,|\xi|)=\frac{1}{4}\lambda\left(|\Phi|^{2}-\eta^{2}\right)^{2}+
\frac{1}{4}\lambda_{\xi}\left(|\xi|^{2}-\eta_{\xi}^{2}\right)^{2} +
\beta|\Phi|^{2} |\xi|^{2},
\end{equation}
where the complex scalar field $\xi$ represents the bosonic supercurrent.
To form a stable string imposes the condition
\begin{equation}
\label{eq-nocurrent}
\lambda \eta^{4} > \lambda_{\xi}\eta_{\xi}^{4}.
\end{equation}
To prevent the supercurrent from becoming non-zero outside the string core
imposes an additional constraint
\begin{equation}
\label{eq-conone}
\frac{2\beta}{\lambda_{\xi}}\geq\frac{\eta^{2}_{\xi}}{\eta^{2}}.
\end{equation}
At the center of the homilia $|\Phi|=\kappa(e)\eta\neq0$, therefore
a non-zero current can only exist at the homilia string core for
\begin{equation}
\label{eq-contwo}
\frac{2\beta}{\lambda_{\xi}}\leq\frac{\eta^{2}_{\xi}}{\kappa(e)^{2}\eta^{2}},
\end{equation}
Combining Eqs.\ (\ref{eq-conone}) and (\ref{eq-contwo}) yields
\begin{equation}
\label{eq-boson}
1\leq \frac{2\beta\eta^{2}}{\lambda_{\xi}\eta_{\xi}^{2}}\leq \frac{1}{\kappa(e)^{2}}.
\end{equation}
Depending on the choice of $e$, Eq.\ (\ref{eq-boson}) indicates
that homilia strings will only support bosonic
supercurrents for a narrow range of parameters, and in general homilia
strings will not support bosonic superconductivity. Furthermore, the
vorton current is unstable and will drift off the string core unless \cite{davis88b}
\begin{equation}
\label{eq-drift}
\frac{{\cal N}^{2}}{R^{2}} \leq 2\beta\eta^{2}-\lambda_{\xi}\eta_{\xi}^{2},
\end{equation}
where ${\cal N}$ is a constant that represents the current phase winding number
and $R$ is the radius of the vorton. From Eq.\ (\ref{eq-boson}) and
Eq.\ (\ref{eq-drift}) we find that
\begin{equation}
\label{eq-compare}
\frac{{\cal N}^{2}}{R^{2}} \leq \left(2\beta\eta^{2}-\lambda_{\xi}\eta_{\xi}^{2}\right)\leq (\kappa(e)^{-2}-1)\lambda_{\xi}\eta^{2}_{\xi}.
\end{equation}
The quantity ${\cal N}^{2}/R^{2}$ can be determined \cite{davis88b}, to give
\begin{equation}
\label{eq-ratio}
\frac{{\cal N}^{2}}{R^{2}} \approx \frac{\mu}{\pi\delta^{2}\eta_{\xi}^{2}} \sim \frac{\lambda\eta^{4}}{\pi\eta_{\xi}^{2}},
\end{equation}
where $\delta$ is the unperturbed string width. Combining Eq.\ (\ref{eq-ratio})
with Eq.\ (\ref{eq-compare}) results in
\begin{equation}
\lambda\eta^{4} \leq (\kappa(e)^{-2}-1) \pi \lambda_{\xi}\eta_{\xi}^{4},
\end{equation}
which clearly contradicts Eq.\ (\ref{eq-nocurrent}) when $\kappa(e)^{-2}-1<\pi^{-1}$.
Since there is an empirical bound on $e$, in the
electroweak model, given by $e<2$ (Sec.\ \ref{sec-vortex}) we write
\begin{equation}
\kappa(e) > \kappa(e=2) \approx 0.96,
\end{equation}
and stable bosonic superconducting electroweak homilia vortons are ruled out.

To describe a homilia vorton with a fermonic current, we 
write the field for a homilia string (e.g., an $\alpha$-string) as
\begin{equation}
\Phi^{\alpha}=e^{i\tau^{\alpha}\theta}\left(\begin{array}{c} f^{\alpha}(r) \\ f^{\beta}(r)\end{array}\right),
\end{equation}
where the phase of the $\chi^{\beta}$-field has been set to zero. 
Fermionic currents are coupled to the $\Phi$-field through the
Yukawa interaction. If we include neutrinos and
neglect gauge fields, the interaction Lagrangian is
\begin{eqnarray}
\label{eq-fermionic}
{\cal L}_{F} & = & i\overline{\psi}_{\nu}\gamma^{\mu}\partial_{\mu}\psi_{\nu}+i\overline{\psi}_{L}\gamma^{\mu}\partial_{\mu}\psi_{L}+i\overline{\psi}_{R}\gamma^{\mu}\partial_{\mu}\psi_{R}
-g\overline{\psi}_{L}\phi^{2}\psi_{R} \nonumber \\
& & -g\overline{\psi}_{R}\overline{\phi^{2}}\psi_{L}-g\overline{\psi}_{\nu}\phi^{1}\psi_{R}-g\overline{\psi}_{R}\overline{\phi^{1}}\psi_{\nu},
\end{eqnarray}
where $\psi_{\nu}$ and $\psi_{L}$ are components of the left-handed isodoublet,
$\psi_{R}$ is the right-handed isosinglet and $g$ denotes the coupling constant.
To incorporate the homilia string into Eq.\ (\ref{eq-fermionic}), we must rotate
the $\Phi$-field so that $\Phi^{\alpha}(r\rightarrow\zeta/2)\rightarrow\Phi_{EW}$.
Since we have neglected gauge fields the rotation is achieved using the global
gauge transformation $V$ in Eq.\ (\ref{eq-ewtrans}).
This transform results in
\begin{equation}
\Phi^{\alpha}\rightarrow V\Phi^{\alpha} =\left(\begin{array}{c} -f^{\alpha}(r)e^{i\theta}+f^{\beta}(r) \\ f^{\alpha}(r)e^{i\theta}+f^{\beta}(r) \end{array}\right)=\left(\begin{array}{c} \phi^{1}(r,\theta) \\ \phi^{2}(r,\theta) \end{array}\right).
\end{equation}
After performing the gauge transformation, we can substitute the forms of
$\phi^{1}$ and $\phi^{2}$ into the equations of motion for the lepton fields.
The equations of motion are
\begin{equation}
\label{eq-fsc}
i\gamma^{\mu}\partial_{\mu}\left(\begin{array}{c} \psi_{L} \\ \psi_{R} \\ \psi_{\nu} \end{array}\right) =
\left(\begin{array}{ccc} 0 & \phi^{2} & 0 \\ \overline{\phi^{2}} & 0 & \overline{\phi^{1}} \\ 0 & \phi^{1} & 0 \end{array} \right)
\left(\begin{array}{c}\psi_{L} \\ \psi_{R} \\ \psi_{\nu} \end{array}\right).
\end{equation}
Because the $\Phi$-field rotates in isospin space inside the homilia string
core (i.e., $\phi^{1}$ and $\phi^{2}$ are non-zero), the fields 
$\psi_{L}$, $\psi_{R}$ and $\psi_{\nu}$ all become coupled. We also note that
in general $\phi^{1}\neq\phi^{2}$.

From Eq.\ (\ref{eq-fsc}), we derive the relationship
\begin{equation}
\label{eq-psirel}
\frac{\gamma^{\mu}\partial_{\mu} \psi_{L}}{\phi^{2}}=\frac{\gamma^{\mu}\partial_{\mu} \psi_{\nu}}{\phi^{1}}.
\end{equation}
Multiplying both sides of Eq.\ (\ref{eq-psirel}) by $\gamma^{1}$ we obtain
\begin{equation}
\label{eq-psirelx}
\frac{\partial_{x} \psi_{L}}{\phi^{2}}=\frac{\partial_{x} \psi_{\nu}}{\phi^{1}}.
\end{equation}
Equation (\ref{eq-psirelx}) has the solution
\begin{mathletters}
\label{eq-psisoln}
\begin{eqnarray}
\psi_{L} & = & \tau \int C_{1} B(t,x,y,z) \phi^{2}\,dx \\
\psi_{\nu} & =& \tau \int C_{1} B(t,x,y,z) \phi^{1}\,dx,
\end{eqnarray}
\end{mathletters}
where $C_{1}$ is a constant, $\tau$ describes the spinor properties of 
$\psi_{L}$ and $\psi_{\nu}$ and $B(r,\theta)$ is a non-zero function.
However, if we multiply Eq.\ (\ref{eq-psirel}) by $\gamma^{2}$ we obtain
\begin{equation}
\label{eq-psirely}
\frac{\partial_{y} \psi_{L}}{\phi^{2}}=\frac{\partial_{y} \psi_{\nu}}{\phi^{1}}.
\end{equation}
Substituting Eqs.\ (\ref{eq-psisoln}) into Eq.\ (\ref{eq-psirely}) results
in the condition
\begin{equation}
\label{eq-condition}
C_{1} \phi^{1} \partial_{y} \int B(t,x,y,z) \phi^{2}\,dx = C_{1} \phi^{2} \partial_{y} \int B(t,x,y,z) \phi^{1}\,dx.
\end{equation}
To solve Eq.\ (\ref{eq-psirely}) we utilize the fact that
$\phi^{1}$ and $\phi^{2}$ are non-zero functions, determined by the vortex 
solution. Since $\phi^{1}\neq\phi^{2}$, the solution
to Eq.\ (\ref{eq-condition}) requires (for $C_{1}\neq0$)
\begin{mathletters}
\begin{eqnarray}
\int \phi^{1} dy & = & \int B(t,x,y,z) \phi^{1} dx \\
\int \phi^{2} dy & = & \int B(t,x,y,z) \phi^{2} dx,
\end{eqnarray}
\end{mathletters}
which cannot be satisfied in general.
Therefore, Eq.\ (\ref{eq-condition}) requires $C_{1}=0$ and this results
in the solutions for the lepton fields having zero values, i.e., $\psi_{L}=\psi_{R}=\psi_{\nu}=0$. Homilia strings
cannot support fermonic supercurrents which is a direct consequence of the
rotation of the $\Phi$-field in isospin space, which leads to the coupling of the
fermionic current to neutrino fields.

Although homilia strings are expected to support
non-Abelian gauge supercurrents \cite{everett88}, such supercurrents
are unlikely to support vorton states.
This is because vorton stability is associated with conserved quantum numbers
related to the Noether charge and to current phase winding in the loop \cite{davis88}.
Such quantum numbers do not appear to exist for non-Abelian gauge
supercurrents \cite{kibble97}, and hence non-Abelian gauge
superconducting vortons are unlikely.
Since homilia string loops do not support fermionic superconductivity
or bosonic vortons, they do not settle into vorton states.
Hence homilia strings will avoid the adverse
cosmological consequences of stable loops.

\subsection{GUT scale homilia strings}
If symmetry breaking of an $SU(N)\times U(1)$ model occurs at the GUT phase transition,
GUT scale homilia strings will also form. In such models
there are $N$ types of homilia strings, which after performing an appropriate  
gauge transformation are wound in the $N$ complex components of the $\Phi$ 
multiplet. In a network of homilia
strings, consisting of all $N$ types of homilia strings, the network texture
is topologically stable. Due to the number of possible gauge fields, the
vortex solution of such a configuration depends on the order of the symmetry
breaking group. Simulations of global homilia strings in $SU(3)$ symmetry
breaking indicate that they are stable against collapse into $SU(2)$ global homilia
strings. This suggests that GUT homilia strings would be stable.
For symmetry groups of the form $SU(N)\times U(1)$, the formation of sphalerons
requires the intersection of all $N$ types of homilia strings at a point.
Such an intersection is unlikely for $N\geq 3$ and hence
we do not expect GUT sphalerons to form.

\subsection{Network evolution}
\label{sec-network}
GUT homilia strings retain the same useful cosmological properties as cosmic strings
GUT models which do not predict the formation of
stable cosmic string defects can nevertheless accommodate large scale structure and
density fluctuations induced by GUT homilia strings. Since homilia
strings have a mass per unit length,
cylindrical symmetry and are Lorentz invariant along the string length,
they will also exhibit a conical spacetime similar to cosmic strings \cite{hiscock85}.
Given a GUT phase transition at an appropriate scale,
it is therefore possible for homilia strings to produce large scale structure
through `wakes' \cite{vachaspati91,perivolaropoulos90} and collapsing string
loops \cite{scherrer89}.

The evolution of a homilia string network can be described in a manner identical
to $U(1)$ cosmic string networks \cite{shellard87}.
Loops arise from intercommuting homilia strings,
with subsequent collapse and annihilation providing an energy loss mechanism
for the network \cite{shellard87}. In this way, an homilia string network is
prevented from dominating the energy density of the Universe.
Since homilia strings only intercommute with strings of the same type
(Sec.\ \ref{sec-force}), we can model an
homilia string network consisting of different string types as
the superposition of two (or more) independent networks. We write
\begin{equation}
\Gamma_{\infty}(t) = N \rho_{\infty}(t),
\end{equation}
where $\Gamma_{\infty}$ is the total energy density of infinitely long homilia 
strings, $N$ is the order of the symmetry breaking group, which leads to
$N$ types of homilia strings, and $\rho_{\infty}$ is the energy
density of the strings for an independent network of strings of the
same type. The energy density of infinite strings in a standard cosmic string network
would also be described by $\rho_{\infty}$,
which for large $t$ approaches the scaling solution $\rho_{\infty}\sim t^{-2}$.
Comparing the characteristic length scale of homilia strings, $L(t)$, in the 
network to that of cosmic strings we find
\begin{equation}
\label{eq-lensol}
L(t)=\sqrt{\frac{\mu_{H}}{\Gamma_{\infty}(t)}}=\frac{1}{\sqrt{N}} L_{CS}(t),
\end{equation}
where $\mu_{H}$ is the energy per unit length of a homilia string
and $L_{CS}(t)$ is the characteristic length scale of a cosmic string
network. Because homilia strings are inefficient at intercommuting and loop formation
(i.e., they only intercommute with strings of the same type),
the characteristic length scale of strings in an homilia string network is smaller than that
for strings in a cosmic string network, at a given time $t$. Equation (\ref{eq-lensol})
indicates that the number density of a network of
homilia strings will be greater than the corresponding value for a cosmic string network.

\subsection{Cosmic Microwave Background fluctuations}
The evolution of an homilia string network leaves characteristic
fluctuations in the cosmic microwave background (CMB), providing an experimental
test for homilia strings. Since homilia strings possess a conical metric, GUT
homilia strings will be capable of inducing non-gaussian fluctuations in the CMB
in the same manner as cosmic strings \cite{kaiser84}.

The distance between homilia strings, $L(t)$, is proportional to $1/\sqrt{N}$
[see Eq.\ (\ref{eq-lensol})], hence CMB
fluctuations induced by homilia strings will have a smaller angular size
than for cosmic strings. Based on estimates of cosmic string fluctuations
\cite{bouchet88}, the angular size of the fluctuations induced by homilia
strings, at the time of last scattering, is given approximately by
\begin{equation}
\label{eq-cmb}
\theta\sim \frac{1}{3 \sqrt{N}} z_{ls}^{-1/2}\,\mbox{rad}\approx \frac{0.6^{o}}{\sqrt{N}},
\end{equation}
where $z_{ls}$ is the redshift at the surface of last scattering
($z_{ls}\sim10^{3}$). Hence we would expect a GUT scale homilia
string network to provide much greater power in the angular power spectrum 
of the CMB anisotropy (at small angular scales), than a network of cosmic strings.
Measurements of the anisotropy angular power spectrum may provide a means of differentiating fluctuations due to
homilia strings from those induced by cosmic strings. However, to quantify the magnitude of these
fluctuations will require a detailed simulation of an
homilia string network.

\section{Conclusion}
\label{sec-conclusion}
In this paper we have introduced a new type of topological string
defect, called a homilia string, which arises in
$SU(N)\times U(1)$ models of symmetry breaking. Homilia strings are classically
stable in $SU(2)\times U(1)$, consequently, they
are a prediction of the standard electroweak unification model.
Individual homilia strings are stabilized via their boundary conditions with adjacent defects.
The topological stability of these string defects is a consequence of
the surrounding homilia string network, which can be considered to be a texture.
Although some aspects of homilia string dynamics are similar
to $U(1)$ cosmic strings, homilia strings have
additional properties which may have important consequences for cosmology.
Sphalerons in $SU(2)\times U(1)$ are identified with the intersection of two
homilia strings. Due to repulsive forces, homilia strings separate and
induce sphaleron decay. Electroweak homilia strings loops do not support bosonic or
fermionic vorton states and hence avoid the adverse consequences
associated with vortons. Homilia strings 
retain the advantages of GUT cosmic strings for inducing large scale
structure in the observable Universe and non-gaussian fluctuations
in the cosmic microwave background. The anisotropy angular power spectrum of homilia
strings is expected to exhibit greater power, at small angular scales, than
the spectrum originating from cosmic strings.

Future work will be directed at examining the detailed cosmological
ramifications of homilia strings, and their potential for experimental
verification. Investigating
homilia strings in more complex GUTs, which support cosmic strings
(e.g., $SU(2)\times U(1)\times Z_{2}$), would also allow us to study
the interactions between homilia strings and cosmic strings. To understand
in detail the differences between homilia strings and cosmic strings
will require detailed simulations of large scale homilia string networks.

\acknowledgments
One of the Authors (MJT) acknowledges the support of an APA scholarship.

\end{document}